\documentclass[aps,prl,onecolumn,groupedaddress]{revtex4}
\usepackage{graphicx}
\begin{document}
\title{Heralded generation of entangled photon pairs}
\author{Stefanie Barz$^{1,2,\dagger}$, Gunther Cronenberg$^{1,2 \dagger}$, Anton Zeilinger$^{1,2}$, Philip Walther$^{1,2}$}
 \affiliation{$^1$~Faculty of Physics, University of Vienna, Boltzmanngasse 5, A-1090 Vienna, Austria\\
 $^2$~Institute for Quantum Optics and Quantum Information (IQOQI), Austrian Academy of Sciences, Boltzmanngasse 3, A-1090 Vienna, Austria\\
 $^{\dagger}$~These authors contributed equally to this work}
\begin{abstract}
Entangled photons are a crucial resource for quantum communication and linear optical quantum computation. Unfortunately, the applicability of many
photon-based schemes is limited due to the stochastic character of the  photon sources. Therefore, a worldwide effort has focused in
overcoming the limitation of probabilistic emission by generating two-photon entangled states conditioned on the detection of auxiliary photons.
Here we present the first heralded generation of photon states that are maximally entangled in polarization with linear optics and standard photon
detection from spontaneous parametric down-conversion \cite{Kwiat1995}. We utilize the down-conversion state corresponding to the generation of three photon pairs,
 where the coincident detection of four auxiliary photons unambiguously heralds the successful preparation of the entangled state \cite{Sliwa2003}.
 This controlled generation of entangled photon states is a significant step towards the applicability of a linear optics quantum network, in particular for
 entanglement swapping, quantum teleportation, quantum cryptography and scalable approaches towards photonics-based quantum computing \cite{Nielsen2000}.
\end{abstract}

\maketitle

Photons are generally accepted as the best candidate for quantum communication due to their lack of decoherence and their possibility of being easily manipulated. However, it has also been discovered that a scalable quantum computer can in principle be realized by using only single-photon sources, linear optical elements and single-photon detectors \cite{Knill2001a}. Several proof-of-principle demonstrations for linear optical quantum computing have been given, including controlled-NOT gates \cite{ Pittman2001, OBrien2003, Pittman2003, Gasparoni2004}, Grover's search algorithm \cite{Kwiat2000a,Prevedel2007a}, Deutsch-Josza algorithm \cite{Tame2007}, Shor's factorization algorithm \cite{Lu2007,Lanyon2007} and the promising model of the one-way quantum computation \cite{Walther2005a}.

A main issue on the path of photonic quantum information processing is that the best current source for photonic entanglement, spontaneous parametric down-conversion (SPDC), is a process where the photons are created at random times. All photons involved in a protocol need to be measured including
a detection of the desired output state. This impedes the applicability of many of the beautiful proof-of-principle experiments, especially when dealing with multiple photon pairs \cite{Nielsen2000} and standard detectors without photon number resolution.
Other leading technologies in this effort are based on other physical systems including single trapped atoms and atomic ensembles \cite{Kimble2008}, quantum dots \cite{Michler2000}, or nitrogen-vacancy centers in diamond \cite{Kurtsiefer2000}.
Although these systems are very promising candidates, each of these quantum state emitters faces significant challenges for realizing heralded entangled states; typically due to low coupling efficiencies, the uncertainty in emission time or the distinguishability in frequency of the photons created.

However, the probabilistic nature originating from SPDC can be overcome by several approaches conditioned on the detection of auxiliary photons \cite{Sliwa2003,Pittman2003b,Walther2007}. It was shown that the production of one heralded polarization-entangled photon pair using only conventional down-conversion sources, linear optical elements, and projective measurements requires at least three entangled pairs \cite{Kok2000a}. Here we describe an experimental realization for producing heralded two-photon entanglement along theses lines, suggested by \'{S}liwa and Banaszek that relies on triple-pair emission from a single down-conversion source \cite{Sliwa2003}.
This scheme shows significant advantages compared to other schemes where either several SPDC sources and two-qubit logic gates \cite{Pittman2003b} or more ancilla photons \cite{Walther2007} are required.

\begin{figure}[b]
\includegraphics[width=0.80\textwidth]{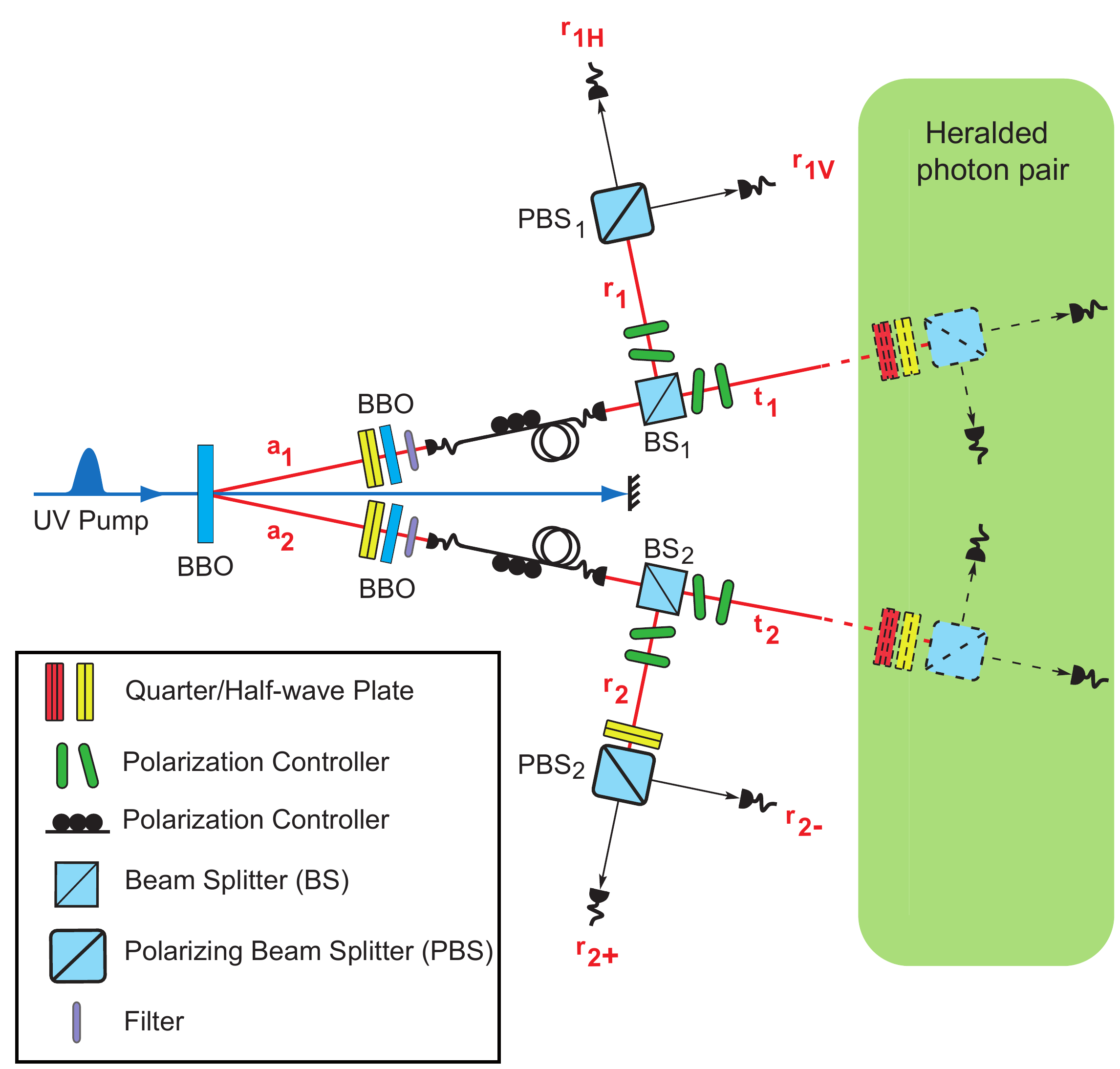}
\caption{\label{figure1}Setup for the heralded generation of entangled photon pairs. Six photons are created simultaneously
by exploiting higher-order emissions in a spontaneous parametric down-conversion process. The photons are passing a
narrowband filter and are coupled to single-mode fibers. They are brought to beam splitters and the reflected modes are
analyzed in $\left|H/V\right\rangle$ basis and in $\left|\pm\right\rangle$ basis, respectively, using polarizing
beam splitters (PBS) and a half-wave plate (HWP) orientated at $45^\circ$. State characterization of the heralded
photon pair in the transmitted modes is performed via polarization analysis and the help of quarter-wave plates (QWPs), HWPs and PBSs.}
\end{figure}

Current down-conversion experiments allow for the simultaneous generation of three photon pairs  \cite{Zhang2006, Wieczorek2009, Prevedel2009, Radmark2009} with typical detection count rates, dependent on the experimental configuration, of about $10^{-3}$ to $10^{-1}\,$Hz. We use a setup of this kind such that the coincident detection of four auxiliary photons is used to predict the presence of two polarization-entangled photons in the output modes. The auxiliary photons thus herald the presence of a Bell state and it is not necessary to perform a measurement on that state to confirm its
presence.

Figure \ref{figure1} gives a schematic diagram of our setup to generate the heralded state $\left|\phi^+\right\rangle=\frac{1}{\sqrt{2}}\left(\left|H\right\rangle_{t_1}\left|H\right\rangle_{t_2}+\left|V\right\rangle_{t_1}\left|V\right\rangle_{t_2}\right)$, where $H$ and $V$ denote horizontal and vertical polarization, respectively,  whereas $t_1$ and $t_2$ correspond to the transmitted modes after the beam splitters. For generating the heralded state, $\left|\phi^+\right\rangle$, three photon pairs have to be emitted simultaneously into spatial modes
$a_{1}$ and $a_{2}$, resulting in:
\begin{eqnarray*}
\left|\Psi_3\right\rangle =1/2\cdot (\left|HHH\right\rangle_{a_1}\left|VVV\right\rangle_{a_2}-\left|HHV\right\rangle_{a_1}\left|VVH\right\rangle_{a_2}
+\left|VVH\right\rangle_{a_1}\left|HHV\right\rangle_{a_2}-\left|VVV\right\rangle_{a_1}\left|HHH\right\rangle_{a_2})
\end{eqnarray*}

These photons are guided to non-polarizing beam splitters ($\textrm{BS}_1$ and $\textrm{BS}_2$) with various splitting ratios. Our scheme only succeeds when four photons are reflected and measured in a four-fold coincidence. The two reflected photons of $\textrm{BS}_1$ are projected onto $\left|H/V\right\rangle$ basis for mode $r_{1}$, 
while the two reflected photons of $\textrm{BS}_2$ are measured in $\left|\pm\right\rangle=\frac{1}{\sqrt{2}}(\left|H\right\rangle\pm\left|V\right\rangle)$ basis for mode $r_{2}$.
We are interested in the case where one photon is present in each of the modes $r_{1H,1V}$ and $r_{2+,2-}$. Considering only these terms, the output state results in
\begin{eqnarray}
	\left|\Psi_3\right\rangle = C(\theta_1,\theta_2)\cdot \left|H\right\rangle_{r_{1H}}\left|V\right\rangle_{r_{1V}}\left|+\right\rangle_{r_{2+}}\left|-\right\rangle_{r_{2-}}
\cdot \frac{1}{\sqrt{2}}\left(\left|H\right\rangle_{t_1}\left|H\right\rangle_{t_2}+\left|V\right\rangle_{t_1}\left|V\right\rangle_{t_2}\right)
\end{eqnarray}
where $C(\theta_1,\theta_2)$ is a constant depending on the transmission coefficients of the beam splitters. The coincident detection of one and only one photon in  the modes $r_{1H}$, $r_{1V}$, $r_{2+}$ and $r_{2-}$ heralds the presence of an entangled photon pair in state $\left|\phi^+\right\rangle$ in the output modes $t_1,t_2$. In the present scheme such a case can only be achieved by three-pair  emission from SPDC. The contribution from two-pair emission is suppressed by destructive quantum interference in the HWP rotation used for $r_{2+,2-}$.
This quantum interference together with the use of number-resolving detectors ensures that the remaining two photons are found in the transmitted modes. If a high transmission of the beam splitters is chosen, it still can be assumed with high probability that the two photons are transmitted even without the use of number-resolving detectors.

In our case of using standard detectors (PerkinElmer photo-avalanche diodes) the transmission of the non-polarizing beam splitters should
ideally be as high as possible such that a measured four-photon coincidence
corresponds to precisely four photons and thus heralds our desired state. Therefore for demonstrating this
dependency we choose beam splitters with different transmission rates of $17\,\%$, $50\,\%$ and $70\,\%$.
Obviously the disadvantage of increasing
the probability of heralding a $\left|\phi^+\right\rangle$ state - which in
principle can be approximately unity - is a reduction in the four-fold
coincidence rate for triggering this state. Only the recent improvements of
laser sources enable stable UV beams with sufficient power for keeping the
measurement time reasonable; in our case the typical counting
 time for one measurement setting varies from 24h to 72h. The actual rate $R$ of the
four-fold coincidences is about $R_{17/83}=83$, $R_{50/50}=14$ and
$R_{70/30}=0.4$ per minute.

\begin{figure}[t]
\includegraphics[width=0.55\textwidth]{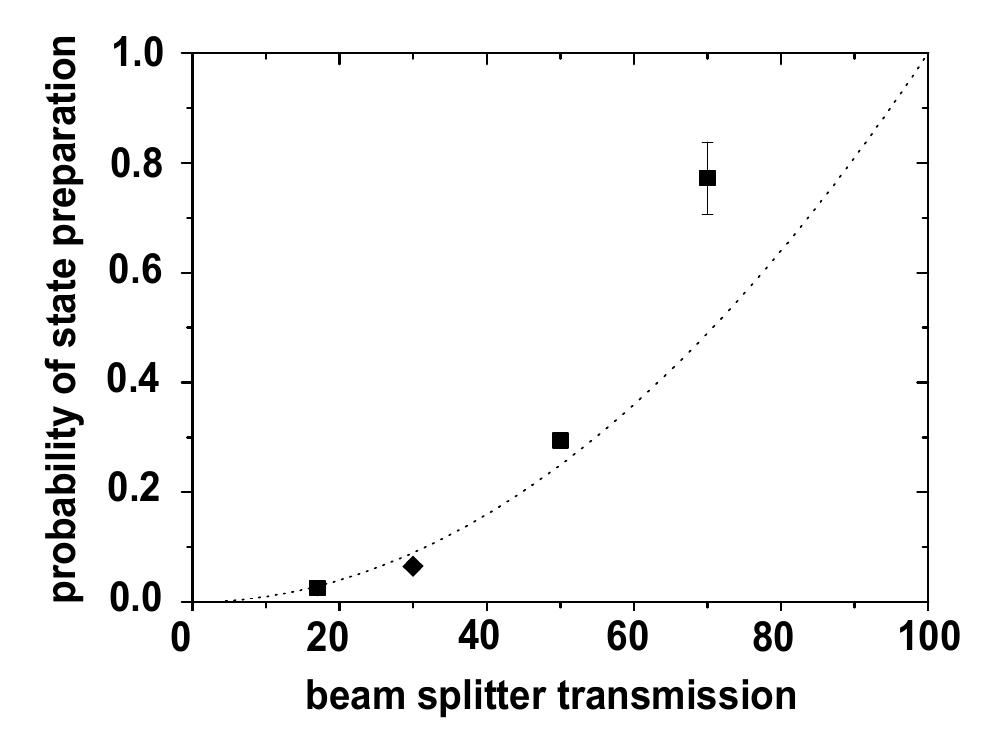}
\caption{\label{figure2}Probability of heralded entanglement generation shown for various beam splitter transmissions. The deviation from the expected
quadratic behavior (line) originates from high-order emissions, which increase the probability of measuring photon pairs in the output modes for higher beam splitter transmissions.
The diamond-shaped data point originates from the experiment with the reduced laser-power and the error bars follow a Poissonian statistic.}
\end{figure}

For characterizing our heralded state, all polarization-dependent
measurement outcomes in the output modes $t_1$ and $t_2$ that are triggered
by the four-fold coincidence in modes $r_{1H}$, $r_{1V}$, $r_{2+}$, and
$r_{2-}$ are analyzed as a function of the beam splitter transmission. The
measured probabilities of finding the various photon numbers in the output
modes are shown in Table \ref{table1} and allow for the reconstruction of the diagonal
elements of the density matrix in the photon number basis
$\left|n_1\right\rangle_{t_1}\left|n_2\right\rangle_{t_2}$, where $n_1$ and $n_2$
are the Fock or photon-number states per spatial mode (see Supplementary
Information). The dependency of the photon-number statistic on the beam splitter
ratio can be clearly seen as the vacuum contribution $P_{0,0}$
decreases with
higher transmission rates. The resulting probability of heralded entanglement generation is graphically shown in Figure
\ref{figure2} as a function of the beam splitter transmission. These
probabilities, defined as $P=C_6/(C_4\cdot\eta^2)$ where $C_6$ ($C_4$) is
the six(four)-fold coincidence rate and $\eta$ is the total photon detection
efficiency per mode, are  $P_{17/83}=(2.5\pm0.2)\,\%$,
$P_{50/50}=(29.4\pm1.0)\,\%$ and $P_{70/30}=(77.2\pm6.6)\,\%$ for the
different transmission rates. Remarkably, these probabilities are achieved
due to the high visibility of $(86.2 \pm 0.7)\,\%$ for the destructive
four-photo quantum interference.

\begin{table}
	\begin{center}\footnotesize
		\begin{tabular}{l|c|c|c|c}\hline
		                      &17/83                    					 &30/70                       							    &50/50                           &70/30\\\hline
		 $P_{0;0}\ \ $					&$\ \ (9.74\pm0.002)\cdot10^{-1}\ \	$&$\ \ (9.63\pm0.003)\cdot10^{-1}\ \ $ 	  	&$\ \ (9.15\pm0.006)\cdot10^{-1}\ \ $	&$\ \ (8.68\pm0.02)\cdot10^{-1}\ \ $\\
		 $P_{1;0}+P_{0;1}$			&$		(2.57\pm0.02)\cdot10^{-2} 		$&$    (3.67\pm0.03)\cdot10^{-2} $				  &$(8.19\pm0.06)\cdot10^{-2} $		 		  &$(1.23\pm0.03)\cdot10^{-1}$   		 \\
		 $P_{1;1}$							&$		(2.58\pm0.16)\cdot10^{-4}			$&$    (6.14\pm0.35)\cdot10^{-4}$						&$(3.06\pm0.13)\cdot10^{-3}$ 	 			  &$(8.03\pm0.65)\cdot10^{-3}$ 				\\
		 $P_{2;0}+P_{0;2}$			&$		(2.75\pm0.51)\cdot10^{-5}  		$&$    (3.57\pm0.84)\cdot10^{-5}$						&$(3.72\pm0.36)\cdot10^{-4}$		 			 &$(7.49\pm0.14)\cdot10^{-4}$ 				\\
		 $P_{2;1}+P_{1;2}$			&$0																	$&$    (5.94\pm3.43)\cdot10^{-6}$						&$(3.66\pm1.38)\cdot10^{-5}$					 &$(1.07\pm0.76)\cdot10^{-4}$ 				\\
		 $P_{2;2}$							&$0																	$&$0$															          &$0$ 											            &$0$ 												\\\hline
		
		\end{tabular}
	\end{center}
	\caption{List of the photon-number probabilities $P_{n_1;n_2}$ of having $n_1$ and $n_2$ photons in the output modes $t_1$ and $t_2$. The events with the same sum of photon
	numbers in modes $t_1$ and $t_2$ are compared to the probability of obtaining one photon in each of the output modes. The error for each probability is following a Poissonian distribution of the      	 measured counts. See Supplementary Information for a more detailed analysis including polarization modes.}
		\label{table1}
\end{table}

The detection of more than one photon per spatial mode results from eight-
or more-photon emissions due to the technical limitation of working in the
high laser-power regime for optimizing count rates. Obviously, the detected
six-fold coincidences for obtaining the $P_{1,1}$ contribution also capture
the higher-order emission, which cover about $10\,\%$ of the coincidences
for the high laser-power case. Naturally, the averaging of correlated and
anti- or non-correlated measurement results decreases the quantum
correlations of the desired output state. For fair and sensitive
representation of this effect an additional set of polarization-correlation
measurements is performed. Although these polarization measurements are
triggered by a four-fold coincidence, the requirement of obtaining an
additional two-fold coincidence intrinsically leads to a post-selection of
the two-photon polarization density matrix (see Figure \ref{figure4}a and Methods).
The corresponding fidelity, $F^{{post}}$ of
this measured photon pair with the corresponding
entangled quantum state $\left|\phi^+\right\rangle$ is
$F^{{post}}_{17/83}=(63.7\pm4.9)\,\%$,
$F^{{post}}_{50/50}=(57.5\pm3.4)\,\%$, and
$F^{{post}}_{70/30}=(61.9\pm7.7)\,\%$ for the different beam splitter
ratios via local unitary transformations.

\begin{figure}
\includegraphics[width=0.55\textwidth]{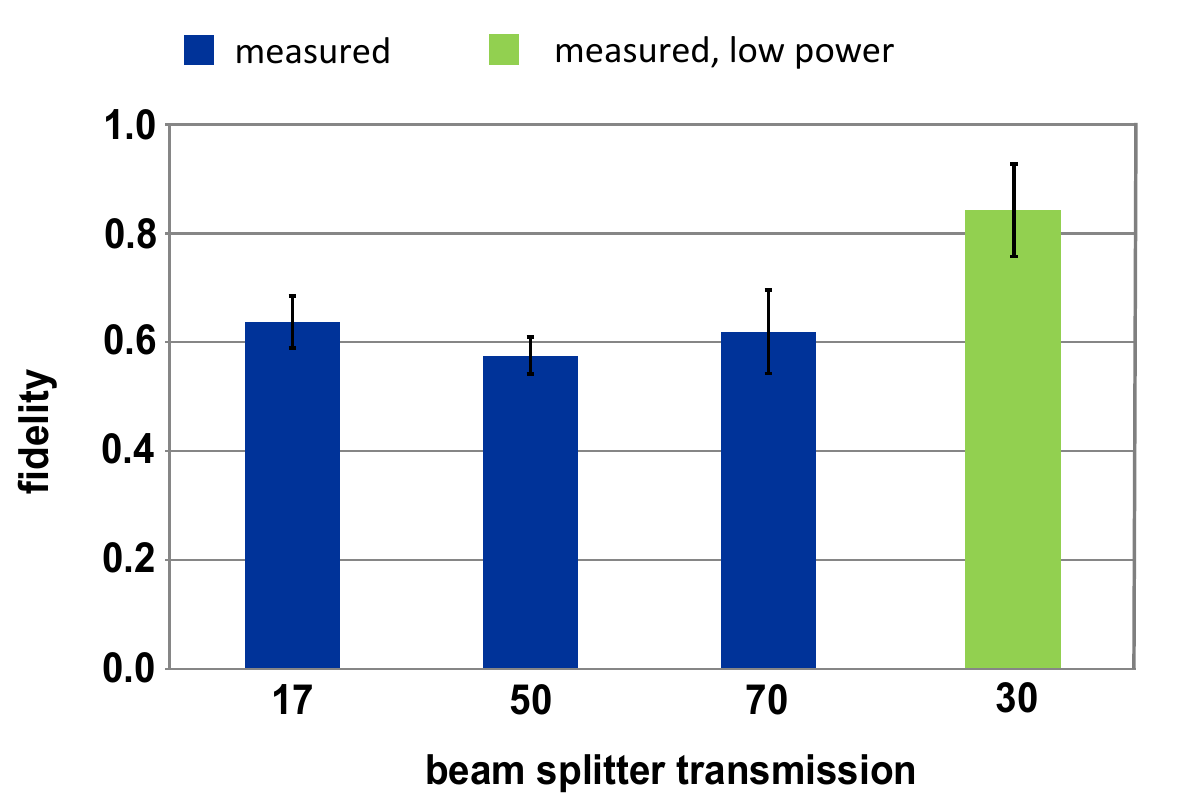}
\caption{\label{figure3}Experimentally obtained fidelities for the two-qubit polarization state with respect to the ideal state $\left|\phi^+\right\rangle$ for various beam splitter transmissions (blue). The effect of higher-order emission is demonstrated by an additional experimental run with a reduced laser-power (green).
For this experiment a beam splitter transmission of $30\,\%$ is chosen for optimizing the required measurement time.
As expected the probability of obtaining the heralded state $\left|\phi^+\right\rangle$
increases with the transmittance of the beam splitters, whereas the polarization-state fidelities are not affected.
The error bars are derived from Monte Carlo simulations based on a Poissonian distribution of the measured counts.}
\end{figure}

\begin{figure}
\includegraphics[width=0.8\textwidth]{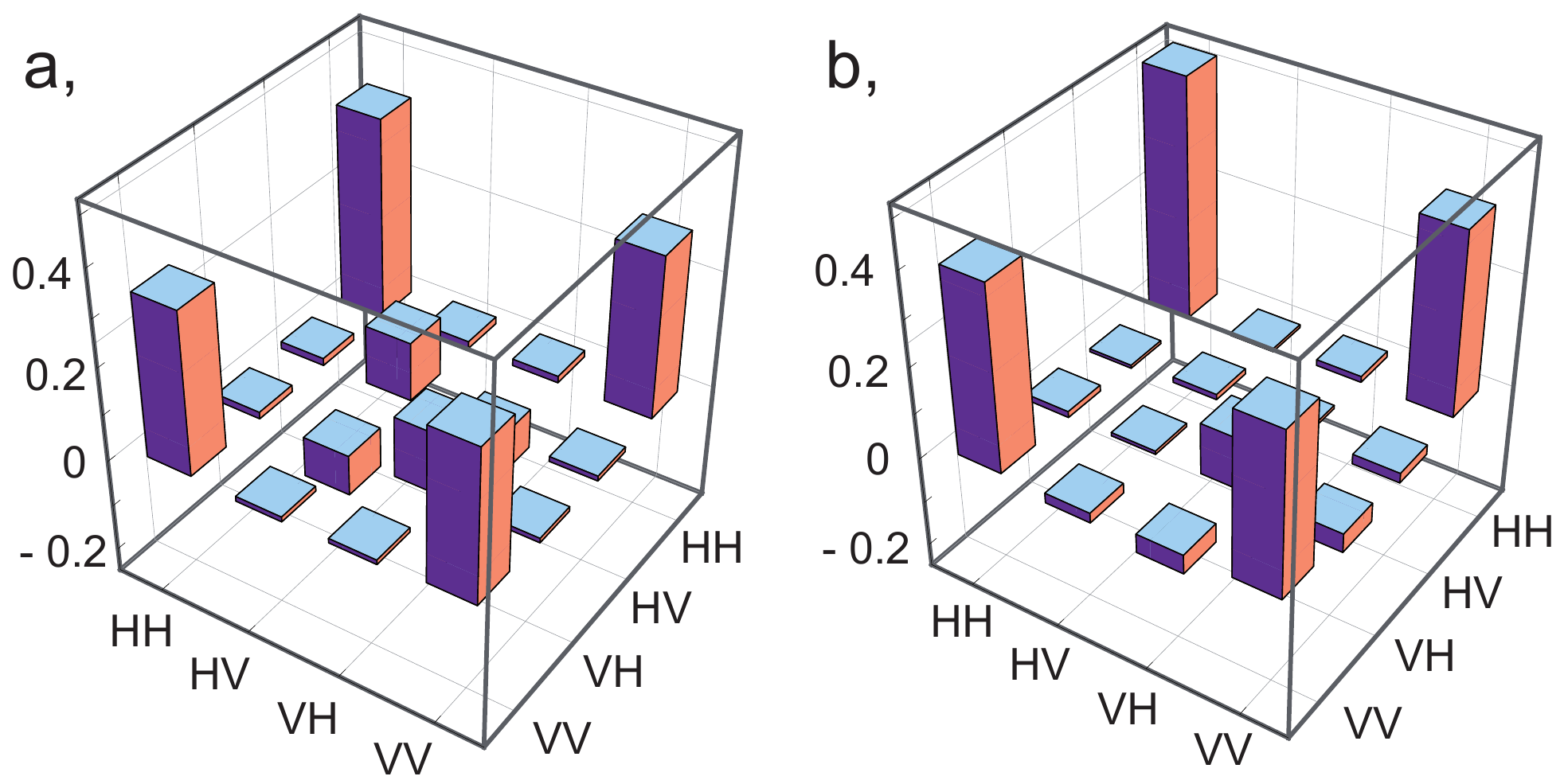}
\caption{\label{figure4}The effect of higher-order emission for the polarization correlations. The trade-off for the increased coincidence rates is manifested in the contribution of higher-order emission. a, Experimentally obtained polarization density matrix with a laser-power of $1.2\,$Watt and a beam splitter transmission of $50\,\%$.
The captured eight-photon contribution leads to a background of a $\left|\psi^-\right\rangle$ state.
b, The reduction of the background is demonstrated when reducing the laser-power. The experimentally reconstructed two-qubit polarization
density matrix is measured with a laser-power of $0.62\,$Watt and a beam splitter transmission of $30\,\%$.
The imaginary part of the density matrices is below $0.09$ for all elements and hence not shown.
}
\end{figure}

For demonstrating the state fidelities' dependency on the laser-power (Figure \ref{figure3}), an
additional experimental run with a reduced laser-power of 620mW and beam
splitter transmissions of
$30\,\%$ was performed. The post-selected density matrix of this state in
Figure \ref{figure4}b clearly shows an improvement of the polarization correlations,
which is quantified
by a fidelity of $F^{{post}}_{30/70}=(84.2\pm8.5)\,\%$. From this data,
we extract the tangle $\tau$ \cite{Coffman2000}, a measure of entanglement
that ranges from 0 for separable states to 1 for maximally entangled states,
as $\tau_{30/70}=0.55\pm0.19$. This density matrix, as commonly written
in the coincidence basis, would allow a violation of local realistic
theories by almost 2 standard deviations as it implies a maximum
Clauser-Horne-Shimony-Holt \cite{Clauser1969, Horodecki1995} Bell parameter
of $S=2.36 \pm 0.22$.
This laser-power dependent noise is therefore
not intrinsic in the setup and is only due to technical limitations, which
can be overcome in future experiments by using photon-number discriminating
detectors or with high-efficient down-conversion sources.

These two-photon density matrices together with the measured photon number
probabilities allow to calculate the state fidelity
$F^{{meas}}=\left\langle \phi^+\right|\rho\left|\phi^+\right\rangle$ of
the output state including vacuum and higher-order terms. This measured
total state fidelity can be extracted as $F^{{meas}}_{17/83}=(0.0164\pm
0.0010)\,\%$, $F^{{meas}}_{30/70}=(0.0517\pm0.0029)\,\%$ ,
$F^{{meas}}_{50/50}=(0.176\pm0.013)\,\%$ and
$F^{{meas}}_{70/30}=(0.497\pm 0.041)\,\%$, which have significantly
improved to standard down-conversion sources. These results suggest that
with future gradual increases to the coincidence rate and the fidelity of
entangled photons, the utilization of this scheme for quantum information processing tasks may not be far out of reach.

This experiment presents the first feasible scheme for the generation of
heralded entangled photon pairs with SPDC and single-photon detectors.  This
conditional method achieves a high preparation efficiency of up to
$~77\,\%$ with measured fidelities of up to $~84\,\%$ for the post-selected
two-photon state. These results successfully underline its potential
applicability for entanglement-based technologies. In conclusion, we
highlight a multi-photon experiment that generates heralded entangled states
as required for long-distance quantum communication and scalable quantum
computing. We note that during the course of the work presented here we learned of a parallel experiment by Wagenknecht et al.\cite{Wagenknecht2010}.

The authors are grateful to R.~Prevedel, X.~Ma, M.~Aspelmeyer, \v{C}.~Brukner and T.~Pittman for discussions and G.~Mondl for
	assistance with the electronics. This work was supported by the Austrian Science Fund (FWF), the Intelligence Advanced Research
	Projects Activity IARPA under Army Research Office ARO, the European Commission under the Integrated
	Project Qubit Applications (QAP) and Quantum Interfaces, Sensors, and Communication based on
	Entanglement (Q-ESSENCE) and the IST directorate, the ERC Senior Grant (QIT4QAD) and the Marie-Curie
	research training network EMALI.
\\

\section{Methods}
We simultaneously produce six photons in the $\left|\Psi_3\right\rangle$-state by using higher-order emissions of a non-collinear type-II SPDC process.
A mode-locked Mira HP Ti:Sa oscillator is pumped by a  Coherent Inc.~Verdi V-18 laser to reach output powers high enough to be able to exploit third-order SPDC emissions.
The pulsed-laser output ($\tau$ = $200\,$fs, $\lambda$= $808\,$nm, $76\,$MHz) is frequency-doubled using a $2\,$mm-thick Lithium triborate (LBO) crystal, resulting in UV pulses of $1.2\,$W cw average.
We achieve a stable source of UV-pulses by translating the LBO with a stepper motor to avoid optical damage to the anti-reflection coating of the crystal (count rate fluctuations less than $3\,\%$ over $24\,$h). Afterwards, dichroic mirrors are used to separate the up-converted light from the infrared laser light. The UV beam is focused on a $2\,$mm-thick $\beta$-barium borate (BBO) crystal cut for non-collinear type-II parametric down-conversion. HWPs and additional BBO crystals compensate walk-off effects and allow the production of any Bell state.
Narrowband interference filters ($\Delta\lambda$ = $3\,$ nm) are used to spatially and spectrally select the down-converted photons which are then coupled into single-mode fibers that guide them to the analyzer setup. There, the photon pairs are directed to non-polarizing beam splitters whose splitting ratios are chosen dependent on the experiment. The reflected modes then are analyzed in $\left|H/V\right\rangle$ basis and in $\left|\pm\right\rangle$ basis.
At this specific angle, where the HWP rotates the polarization by $45^\circ$, any possible four-photon state, emitted into the four modes, $r_{1H,1V}$ and $r_{2+,2-}$, will result only in a three-fold coincidence because of $\left|+-\right\rangle_{r_2}=(\left|HH\right\rangle_{r_2}-\left|VV\right\rangle_{r_2})/\sqrt{2}$. Thus, these two photons will never be split up at the PBS and therefore never contribute to a fourfold coincidence detection. 
The typical photon coupling rates and detector efficiencies for each spatial mode are about $23\,\%$ and $42\,\%$.

Our density matrix is reconstructed by a tomographic set of measurements, where combinations of the single photon projections
$\left|H/V\right\rangle$, $\left|\pm\right\rangle$, and
$\left|R/L\right\rangle=\frac{1}{\sqrt{2}}(\left|H\right\rangle\pm
i\left|V\right\rangle)$, on each of the two photons in modes $t_1$ and $t_2$
are used. The most likely physical density matrix for our 2-qubit
state is extracted using a maximum-likelihood reconstruction \cite{James2001,
Hradil1997, Banaszek1999}. Uncertainties in quantities extracted from these density matrices are calculated using a Monte Carlo routine and assumed Poissonian errors.



\newpage

\section{Supplementary Information}

The analysis of the polarization-dependent photon-number distribution using the Fock basis is shown in Table~\ref{suptable1}. This data represents the diagonal elements of the density matrix,
i.e. the probabilities of having $n$ photons in each mode, starting from the vacuum
contribution up to the case with one photon in each mode.\\

\vspace{2cm}

\begin{table}[h]
	\begin{center}\footnotesize
		\begin{tabular}{l|c|c|c|c}\hline	
		   &17/83 &30/70 &50/50 &70/30 \\\hline
		 $P_{0,0;0,0}\ \ $	&$\ \ (9.74\pm0.002)\cdot10^{-1}\ \	$	&$\ \ (9.63\pm0.003)\cdot10^{-1}\ \ $ 	&$\ \ (9.14\pm0.006)\cdot10^{-1}\ \ $		 	&$\ \ (8.68\pm0.023)\cdot10^{-1}\ \ $		 \\ \hline
		 $P_{1,0;0,0}$			&$		(7.19\pm0.08)\cdot10^{-3} 		$	&$   	(10.8\pm0.15)\cdot10^{-3}$      	&$		(2.42\pm0.04)\cdot10^{-2}$					&$		 (2.70\pm0.12)\cdot10^{-2}$   		\\
		 $P_{0,1;0,0}$			&$		(8.32\pm0.09)\cdot10^{-3}			$ &$		(10.2\pm0.14)\cdot10^{-3}$				&$		(2.26\pm0.03)\cdot10^{-2}$ 					&$		 (3.25\pm0.13)\cdot10^{-2}$  		\\\hline
		 $P_{0,0;1,0}$			&$		(5.77\pm0.07)\cdot10^{-3}			$	&$		(8.41\pm0.13)\cdot10^{-3}$				&$		(1.58\pm0.03)\cdot10^{-2}$ 				 	&$		 (2.72\pm0.12)\cdot10^{-2}$ 			\\
		 $P_{0,0;0,1}$			&$		(4.47\pm0.65)\cdot10^{-3}			$	&$		(7.20\pm0.12)\cdot10^{-3}$			 	&$		(1.93\pm0.03)\cdot10^{-2}$	 		 		&$		 (3.62\pm0.14)\cdot10^{-2}$ 			\\
		 $P_{1,1;0,0}$			&$		(2.08\pm0.44)\cdot10^{-5}			$	&$		(2.38\pm0.69)\cdot10^{-5}$				&$		(2.41\pm0.35)\cdot10^{-4}$			 		&$		 (3.75\pm1.42)\cdot10^{-4}$ 			\\
		 $P_{1,0;1,0}$			&$		(12.05\pm0.85)\cdot10^{-5}		$ &$		(32.3\pm2.02)\cdot10^{-5}$				&$		(10.26\pm0.68)\cdot10^{-4}$ 				&$		 (2.22\pm0.32)\cdot10^{-3}$			\\\hline
		 $P_{1,0;0,1}$			&$		(2.56\pm0.73)\cdot10^{-5}			$ &$		(0.718\pm1.63)\cdot10^{-5}$				&$		(5.94\pm0.64)\cdot10^{-4}$					&$		 (1.01\pm0.33)\cdot10^{-3}$ 			\\
		 $P_{0,1;1,0}$			&$		(4.88\pm0.87)\cdot10^{-5}			$	&$		(6.84\pm1.76)\cdot10^{-5}$			 	&$		(5.22\pm0.56)\cdot10^{-4}$					&$		 (1.83\pm0.32)\cdot10^{-3}$ 			\\
		 $P_{0,1;0,1}$			&$		(6.28\pm0.66)\cdot10^{-5}			$	&$		(21.5\pm1.52)\cdot10^{-5}$				&$		(9.20\pm0.65)\cdot10^{-4}$					&$		 (2.96\pm0.34)\cdot10^{-3}$ 			\\
		 $P_{0,0;1,1}$			&$		(6.63\pm0.25)\cdot10^{-6}			$	&$		(1.19\pm0.49)\cdot10^{-5}$			 	&$		(1.31\pm0.26)\cdot10^{-4}$					&$		 (3.75\pm1.42)\cdot10^{-4}$ 			\\\hline		
		 $P_{1,1;1,0}$			&0																		&$		(1.98\pm1.98)\cdot10^{-6}$				&$		(2.6\pm1.17)\cdot10^{-5}$						 &$		0$ 															\\
		 $P_{1,1;0,1}$			&0																		&$		(3.96\pm2.80)\cdot10^{-6}$				&$		0$ 																	 &$		(1.07\pm0.76)\cdot10^{-4}$			\\
		 $P_{1,0;1,1}$			&0																	&	0																				&$		 (1.05\pm0.74)\cdot10^{-5}$ 					&$		0$ 														  \\
		 $P_{0,1;1,1}$			&0																	&	0                				 								&$		0$																	 &$		0$ 															\\
		 $P_{1,1;1,1}$			&0																	&	0                  											&$		0$																	 &$		0$ 															\\ \hline
		\end{tabular}
	\end{center}
	\caption{The Table lists the measured probabilities $P_{\emph{n}_{1H},\emph{n}_{1V};\emph{n}_{2H},\emph{n}_{2V}}$ of finding $n_{1H}$, $n_{1V}$, $n_{2H}$ and $n_{2V}$ photons numbers in the output
modes $t_{1H}$,$t_{1V}$,$t_{2H}$ and $t_{2V}$ with orthogonal polarizations H and V for the different beam splitter transmissions using our standard photo-avalanche diodes. The error for each probability is following a Poissonian distribution of the measured counts.}
	\label{suptable1}
\end{table}


For each beam splitter transmission ratio we used a (overcomplete) tomographic set of measurements for each basis $\sigma_x$, $\sigma_y$, $\sigma_z$
to reconstruct the polarization correlations of the (high-power) output state. In Table \ref{tab:countrates} we list the measured detection events that were triggered by a four-fold
coincidence of the ancilla photons.\\

\begin{table}[htbp]
  \centering
    \begin{tabular}{ll|rrrrrrrrr}
   \hline

     17/83  &      &$\ \ \sigma_x^{(1)}\sigma_x^{(2)}$    & $\ \ \sigma_y^{(1)}\sigma_y^{(2)}$    & $\ \ \sigma_z^{(1)}\sigma_z^{(2)}$    & $\ \ \sigma_x^{(1)}\sigma_z^{(2)}$    & $\ \ \sigma_x^{(1)}\sigma_y^{(2)}$    & $\ \ \sigma_z^{(1)}\sigma_y^{(2)}$    & $\ \ \sigma_z^{(1)}\sigma_x^{(2)}$    & $\ \ \sigma_y^{(1)}\sigma_x^{(2)}$    & $\ \ \sigma_y^{(1)}\sigma_z^{(2)}$ \\\hline
          &       &       &       &       &       &       &       &       &       &  \\
          & $\left|0,0;0,0\right\rangle\ $  & 124009 & 116831 & 120881 & 118334 & 109941 & 113054 & 93318 & 119896 & 111629 \\
          & $\left|1,0;0,0\right\rangle$ & 990   & 811   & 776   & 995   & 925   & 716   & 636   & 906   & 831 \\
          & $\left|0,1;0,0\right\rangle$ & 871   & 946   & 1114  & 909   & 785   & 1048  & 969   & 1122  & 1014 \\
          & $\left|0,0;1,0\right\rangle$ & 665   & 661   & 668   & 733   & 659   & 610   & 587   & 779   & 727 \\
          & $\left|0,0;0,1\right\rangle$ & 545   & 447   & 531   & 568   & 528   & 481   & 436   & 595   & 591 \\
          & $\left|1,1;0,0\right\rangle$ & 1     & 0     & 2     & 5     & 2     & 5     & 2     & 1     & 4 \\
          & $\left|1,0;1,0\right\rangle$ & 11    & 3     & 8     & 7     & 8     & 10    & 14    & 6     & 13 \\
          & $\left|1,0;0,1\right\rangle$ & 3     & 8     & 7     & 10    & 11    & 2     & 5     & 11    & 2 \\
          & $\left|0,1;1,0\right\rangle$ & 9     & 13    & 4     & 15    & 11    & 7     & 5     & 11    & 9 \\
          & $\left|0,1;0,1\right\rangle$ & 5     & 2     & 6     & 6     & 2     & 9     & 5     & 4     & 10 \\
          & $\left|0,0;1,1\right\rangle$ & 1     & 0     & 1     & 1     & 0     & 0     & 2     & 1     & 1 \\
          & $\left|1,1;1,0\right\rangle$  & 0     & 0     & 0     & 0     & 0     & 0     & 0     & 0     & 0 \\
          & $\left|1,1;0,1\right\rangle$  & 0     & 0     & 0     & 0     & 0     & 0     & 0     & 0     & 0 \\
          & $\left|1,0;1,1\right\rangle$  & 0     & 0     & 0     & 0     & 0     & 0     & 0     & 0     & 0 \\
          & $\left|0,1;1,1\right\rangle$  & 0     & 0     & 0     & 0     & 0     & 0     & 0     & 0     & 0 \\
          & $\left|1,1;1,1\right\rangle$  & 0     & 0     & 0     & 0     & 0     & 0     & 0     & 0     & 0 \\
                  &       &       &       &       &       &       &       &       &       &  \\\hline
      	30/70		&      &$\ \ \sigma_x^{(1)}\sigma_x^{(2)}$    & $\ \ \sigma_y^{(1)}\sigma_y^{(2)}$    & $\ \ \sigma_z^{(1)}\sigma_z^{(2)}$    & $\ \ \sigma_x^{(1)}\sigma_z^{(2)}$    & $\ \ \sigma_x^{(1)}\sigma_y^{(2)}$    & $\ \ \sigma_z^{(1)}\sigma_y^{(2)}$    & $\ \ \sigma_z^{(1)}\sigma_x^{(2)}$    & $\ \ \sigma_y^{(1)}\sigma_x^{(2)}$    & $\ \ \sigma_y^{(1)}\sigma_z^{(2)}$ \\\hline
          &       &       &       &       &       &       &       &       &       &  \\
          & $\left|0,0;0,0\right\rangle$  & 53770 & 59081 & 61607 & 54745 & 55060 & 57606 & 26923 & 57227 & 59855 \\
          & $\left|1,0;0,0\right\rangle$ & 568   & 683   & 677   & 579   & 584   & 644   & 257   & 708   & 752 \\
          & $\left|0,1;0,0\right\rangle$ & 610   & 571   & 674   & 605   & 600   & 587   & 327   & 587   & 605 \\
          & $\left|0,0;1,0\right\rangle$ & 506   & 491   & 551   & 486   & 450   & 457   & 220   & 516   & 570 \\
          & $\left|0,0;0,1\right\rangle$ & 374   & 463   & 399   & 380   & 499   & 442   & 222   & 408   & 449 \\
          & $\left|1,1;0,0\right\rangle$ & 0     & 0     & 1     & 3     & 2     & 1     & 3     & 2     & 0 \\
          & $\left|1,0;1,0\right\rangle$ & 19    & 4     & 17    & 11    & 11    & 7     & 5     & 17    & 13 \\
          & $\left|1,0;0,1\right\rangle$ & 6     & 10    & 3     & 5     & 6     & 14    & 6     & 10    & 8 \\
          & $\left|0,1;1,0\right\rangle$ & 10    & 21    & 4     & 13    & 6     & 9     & 3     & 8     & 5 \\
          & $\left|0,1;0,1\right\rangle$ & 14    & 3     & 8     & 2     & 4     & 8     & 5     & 7     & 8 \\
          & $\left|0,0;1,1\right\rangle$ & 0     & 0     & 1     & 2     & 0     & 0     & 0     & 1     & 2 \\
          & $\left|1,1;1,0\right\rangle$  & 0     & 0     & 1     & 0     & 0     & 0     & 0     & 0     & 0 \\
          & $\left|1,1;0,1\right\rangle$  & 0     & 0     & 0     & 0     & 2     & 0     & 0     & 0     & 0 \\
          & $\left|1,0;1,1\right\rangle$  & 0     & 0     & 0     & 0     & 0     & 0     & 0     & 0     & 0 \\
          & $\left|0,1;1,1\right\rangle$  & 0     & 0     & 0     & 0     & 0     & 0     & 0     & 0     & 0 \\
          & $\left|1,1;1,1\right\rangle$  & 0     & 0     & 0     & 0     & 0     & 0     & 0     & 0     & 0 \\
          &       &       &       &       &       &       &       &       &       &  \\\hline
    \end{tabular}
  \label{tab:addlabel1}
\end{table}

\begin{table}[htbp]
  \centering
     \begin{tabular}{ll|rrrrrrrrr}
   \hline
   50/50  &      &$\ \ \sigma_x^{(1)}\sigma_x^{(2)}$    & $\ \ \sigma_y^{(1)}\sigma_y^{(2)}$    & $\ \ \sigma_z^{(1)}\sigma_z^{(2)}$    & $\ \ \sigma_x^{(1)}\sigma_z^{(2)}$    & $\ \ \sigma_x^{(1)}\sigma_y^{(2)}$    & $\ \ \sigma_z^{(1)}\sigma_y^{(2)}$    & $\ \ \sigma_z^{(1)}\sigma_x^{(2)}$    & $\ \ \sigma_y^{(1)}\sigma_x^{(2)}$    & $\ \ \sigma_y^{(1)}\sigma_z^{(2)}$ \\\hline
          &       &       &       &       &       &       &       &       &       &  \\
          & $\left|0,0;0,0\right\rangle\ $ & 32429 & 15769 & 16779 & 14303 & 14002 & 16309 & 31535 & 17756 & 15859 \\
          & $\left|1,0;0,0\right\rangle$ & 956   & 443   & 390   & 377   & 387   & 376   & 838   & 446   & 405 \\
          & $\left|0,1;0,0\right\rangle$ & 749   & 394   & 482   & 304   & 292   & 443   & 884   & 398   & 374 \\
          & $\left|0,0;1,0\right\rangle$ & 576   & 277   & 208   & 204   & 253   & 314   & 623   & 338   & 224 \\
          & $\left|0,0;0,1\right\rangle$ & 627   & 319   & 433   & 302   & 292   & 308   & 720   & 312   & 381 \\
          & $\left|1,1;0,0\right\rangle$ & 6     & 1     & 9     & 5     & 4     & 5     & 15    & 1     & 0 \\
          & $\left|1,0;1,0\right\rangle$ & 47    & 3     & 15    & 13    & 13    & 16    & 29    & 20    & 14 \\
          & $\left|1,0;0,1\right\rangle$ & 25    & 38    & 9     & 5     & 14    & 15    & 22    & 12    & 8 \\
          & $\left|0,1;1,0\right\rangle$ & 19    & 18    & 8     & 14    & 4     & 8     & 21    & 14    & 7 \\
          & $\left|0,1;0,1\right\rangle$ & 33    & 2     & 11    & 17    & 9     & 11    & 48    & 12    & 11 \\
          & $\left|0,0;1,1\right\rangle$ & 4     & 1     & 4     & 3     & 0     & 0     & 10    & 2     & 1 \\
          & $\left|1,1;1,0\right\rangle$  & 2     & 0     & 0     & 1     & 0     & 0     & 2     & 0     & 0 \\
          & $\left|1,1;0,1\right\rangle$  & 0     & 0     & 0     & 0     & 0     & 0     & 0     & 0     & 0 \\
          & $\left|1,0;1,1\right\rangle$  & 1     & 0     & 0     & 0     & 0     & 0     & 1     & 0     & 0 \\
          & $\left|0,1;1,1\right\rangle$  & 0     & 0     & 0     & 0     & 0     & 0     & 0     & 0     & 0 \\
          & $\left|1,1;1,1\right\rangle$  & 0     & 0     & 0     & 0     & 0     & 0     & 0     & 0     & 0 \\
                    &       &       &       &       &       &       &       &       &       &  \\\hline
       70/30 &      &$\ \ \sigma_x^{(1)}\sigma_x^{(2)}$    & $\ \ \sigma_y^{(1)}\sigma_y^{(2)}$    & $\ \ \sigma_z^{(1)}\sigma_z^{(2)}$    & $\ \ \sigma_x^{(1)}\sigma_z^{(2)}$    & $\ \ \sigma_x^{(1)}\sigma_y^{(2)}$    & $\ \ \sigma_z^{(1)}\sigma_y^{(2)}$    & $\ \ \sigma_z^{(1)}\sigma_x^{(2)}$    & $\ \ \sigma_y^{(1)}\sigma_x^{(2)}$    & $\ \ \sigma_y^{(1)}\sigma_z^{(2)}$ \\\hline
          &       &       &       &       &       &       &       &       &       &  \\
          & $\left|0,0;0,0\right\rangle$  & 833   & 1371  & 1480  & 1478  & 1402  & 3360  & 3630  & 1692  & 972 \\
          & $\left|1,0;0,0\right\rangle$ & 23    & 55    & 36    & 56    & 52    & 94    & 102   & 64    & 22 \\
          & $\left|0,1;0,0\right\rangle$ & 27    & 43    & 79    & 55    & 38    & 136   & 140   & 57    & 33 \\
          & $\left|0,0;1,0\right\rangle$ & 26    & 42    & 48    & 31    & 45    & 110   & 121   & 61    & 25 \\
          & $\left|0,0;0,1\right\rangle$ & 34    & 47    & 66    & 52    & 44    & 147   & 170   & 76    & 41 \\
          & $\left|1,1;0,0\right\rangle$ & 0     & 1     & 1     & 1     & 2     & 2     & 0     & 0     & 0 \\
          & $\left|1,0;1,0\right\rangle$ & 4     & 1     & 3     & 3     & 4     & 6     & 6     & 4     & 4 \\
          & $\left|1,0;0,1\right\rangle$ & 2     & 5     & 4     & 8     & 4     & 10    & 3     & 3     & 0 \\
          & $\left|0,1;1,0\right\rangle$ & 2     & 4     & 2     & 1     & 2     & 11    & 7     & 3     & 4 \\
          & $\left|0,1;0,1\right\rangle$ & 4     & 0     & 7     & 1     & 4     & 12    & 6     & 4     & 2 \\
          & $\left|0,0;1,1\right\rangle$ & 1     & 0     & 0     & 2     & 2     & 0     & 0     & 2     & 0 \\
          & $\left|1,1;1,0\right\rangle$  & 0     & 0     & 0     & 0     & 0     & 0     & 0     & 0     & 0 \\
          & $\left|1,1;0,1\right\rangle$  & 0     & 0     & 0     & 0     & 1     & 0     & 1     & 0     & 0 \\
          & $\left|1,0;1,1\right\rangle$  & 0     & 0     & 0     & 0     & 0     & 0     & 0     & 0     & 0 \\
          & $\left|0,1;1,1\right\rangle$  & 0     & 0     & 0     & 0     & 0     & 0     & 0     & 0     & 0 \\
          & $\left|1,1;1,1\right\rangle$  & 0     & 0     & 0     & 0     & 0     & 0     & 0     & 0     & 0 \\
                    &       &       &       &       &       &       &       &       &       &  \\\hline

    \end{tabular}
     \caption{We show the measured photon-numbers for each polarization and spatial mode, $\left|n_{1H}, n_{1V}, n_{2H}, n_{2V}\right\rangle$, where $n_{1H}$, $n_{1V}$
denote the photon number for the orthogonal polarization states in output mode 1 and $n_{2H}$, $n_{2V}$ denote the orthogonal
polarization states in output mode 2. The settings for the different measurement bases $\sigma_i^{(1)}\sigma_j^{(2)}$ with $i,j=x,y,z$ are adjusted by
phase retarders in front of the polarizing beam splitters.}
\label{tab:countrates}
\end{table}


\end{document}